\documentclass[runningheads]{llncs}
\usepackage{graphicx}

\usepackage{amsmath}
\usepackage{wrapfig}
\usepackage{setspace}

\usepackage{listings}

\usepackage{bookmark}
\usepackage[T1]{fontenc}
\usepackage{latexsym}
\usepackage{color}
\usepackage{todonotes}
\usepackage{epsfig}
\usepackage{picinpar}
\usepackage{amsfonts}
\usepackage{verbatim}
\usepackage{url}
\usepackage{xspace}
\usepackage{multirow}

\usepackage{multicol}

\def\clingcon{{\sc clingcon}\xspace}
\def\ezcsp{{\sc ezcsp}\xspace}
\def \ezsmt{{\sc ezsmt}\xspace}

\def \diff{{\sc cmodels(diff)}\xspace}

\def \diff{{\sc cmodels(diff)}\xspace}
\def \ezsmtPlus{{\sc ezsmt$^+$}\xspace}
\def \smtil{{\sc smt(il)}\xspace}
\def \smtlib{{\sc smt-lib}\xspace}

\def \ezsmtPlus{{\sc ezsmt+}\xspace}
\def \smtil{{\sc smt(il)}\xspace}

\def\cvcFour{{\sc cvc4}\xspace}
\def\zThree{{\sc z3}\xspace}

\def\yices{{\sc yices}\xspace}

\def\cmodels{{\sc cmodels}\xspace}
\def \diff{{\sc cmodels(diff)}\xspace}

\usepackage{todonotes}

\usepackage{amsmath,amssymb}
\usepackage{latexsym}
\usepackage{xspace}
\usepackage{verbatim}
\usepackage{url}
\usepackage{color}
\usepackage{paralist}
\usepackage{booktabs}
\usepackage{alltt}
\usepackage{caption}
\usepackage{enumitem}

\setitemize{noitemsep} 
\setitemize{topsep=0pt} 

\setenumerate{noitemsep} 
\setenumerate{topsep=0pt}

\def\beq{\begin{equation}}
\def\eeq#1{\label{#1}\end{equation}}

\def\ba{\begin{array}}
\def\ea{\end{array}}

\def\<{\langle}
\def\>{\rangle}

\newcommand{\ignore}[1]{}

\def\beq{\begin{equation}}
\def\eeq#1{\label{#1}\end{equation}}

\begin{document}
\title{SMT-based Constraint Answer Set Solver \ezsmtPlus}
%
%\titlerunning{Abbreviated paper title}
% If the paper title is too long for the running head, you can set
% an abbreviated paper title here
%
\author{Da Shen\inst{1}, Yuliya Lierler\inst{2}
  }%

% First names are abbreviated in the running head.
% If there are more than two authors, 'et al.' is used.
%
\institute{University of Maryland, College Park, Maryland, 20742, USA
\email{dashen@terpmail.umd.edu} \and University of Nebraska at Omaha, Omaha, Nebraska, 68182, USA
\email{ylierler@unomaha.edu}
}

\maketitle              % typeset the header of the contribution

\section{Introduction}

Answer set programming (ASP) is a declarative programming paradigm for solving difficult combinatorial search problems~\cite{bre11}.  
%Although ASP gains increasing popularity and is applied in a wide range of fields, it faces challenges. 
Constraint answer set programming (CASP) is a recent development, which integrates ASP with constraint processing. Often, this integration  allows one to tackle a challenge posed by  the \textit{grounding bottleneck}. Originally, systems that process CASP programs rely on combining algorithms/solvers employed in ASP and constraint processing~\cite{geb09,lierbal17}. Lee and Meng; Susman and Lierler; Lierler and Susman \cite{lee13,sus16b,lie17} proposed an alternative approach that  utilizes  satisfiability modulo theory (SMT) solvers~\cite{BarTin-14}  in design of CASP systems.

System \ezsmt~\cite{sus16b} is a representative of an  SMT-based approach for tackling CASP programs. Based on experimental evidence, \ezsmt often outperforms its peers. 
Yet, it has several limitations. For instance,  it is unable to process a large class of logic programs called \textit{non-tight} \cite{fag94}. This restriction does not allow users, for example, to express transitive closure succinctly.
% between cities on a map where city A is connected to city B, and city B is connected to city C, so cities A and C are also connected. 
%Solving \textit{non-tight} programs is important because such relations are crucial in many applications. 
System \ezsmt is also unable to handle optimization statements or enumerate multiple solutions to a problem.
We extend \ezsmt so that the described limitations are eliminated. We call the new system \ezsmtPlus.

\section{Work So far}
\subsection{Theoretical Foundations} 
To process non-tight programs, we utilize the extension of theory originally developed by Niemela \cite{nie08}, where the author characterizes 
solutions of \textit{non-tight} ``normal''
programs in terms of \textit{level rankings}. Lierler and Susman
generalized
these concepts to programs including such commonly used ASP features as choice rules and denials  \cite{lie17}.  
Lierler and Susman also develop a mapping from a logic program to an SMT logic called \smtil such that the models of a constructed \smtil theory are in one-to-one correspondence with answer sets of the program \cite{lie17}. The developed mappings generalize the ones presented by Niemela~\cite{nie08}. Thus, any SMT solver capable of processing \smtil expressions can be used to find answer sets of logic programs. 
Such generalizations allow  our system to tackle \textit{non-tight} programs.

%Niemela introduced strong level rankings and also illustrated how strongly connected components of a dependency graph of a normal program can be used to enhance the transformation from a normal program to an SMT formula~\cite{nie08}. We generalized these results to logic programs considered in our work~\cite{shen18}.
%These generalizations pave the way to extend the \textit{level-ranking}-approach to the case of CASP programs as illustrated by Lierler and Susman \cite{lie17}.

\subsection{SMT-based Answer Set solver \diff}
In this work, we restricts our input to pure answer set programs, which can be either tight or non-tight. The \diff system follows the tradition of answer set solvers such as {\sc assat}~\cite{fan04} and {\cmodels}~\cite{lierphd}. In place of designing specialized search procedures targeting logic programs, these tools compute a program's completion and utilize Satisfiability solvers~\cite{gom08} -- systems for finding satisfying assignments  for propositional  formulas -- for search.
Since not all models of a program's completion are answer sets of a program, both {\sc assat} and {\cmodels} implement specialized procedures (based on loop formulas~\cite{fan04}) to weed out such models. SMT solvers~\cite{BarTin-14} extend Satisfiability solvers. They process  formulas that go beyond propositional logic and may contain, for example, integer linear expressions. The \diff system utilizes this fact and  translates a logic program into an SMT formula so that any model of this formula corresponds to an answer set of the program. It then uses SMT solvers for search. Unlike \cmodels or {\sc assat}, the \diff system does not need an additional step to weed out unwanted models. Also, it utilizes \smtlib  -- a standard input language of SMT solvers~\cite{smt15} -- to interface with these systems. This makes its architecture  open towards new developments in the realm of SMT solving. There is practically no effort involved in incorporating a new SMT system into the \diff implementation.

The theoretical foundation of the \diff system lies on the generalizations of Niemela's ideas described in section 2.1. In this sense, the \diff system is a close relative of an earlier answer set solver~{\sc lp2diff} developed by Janhunen et al.~\cite{jan09}. Yet, {\sc lp2diff} only accepts programs of a very restricted form. For example, neither choice rules nor aggregate expressions are allowed. Solver {\diff} permits such important modeling constructs in its input. Also, unlike {\sc lp2diff}, the \diff system is able to generate multiple solutions.

\begin{figure}[t!]
\begin{center}
\captionsetup{type=figure}
\includegraphics[trim=0cm 0cm 0.8cm 0cm, width=80mm]{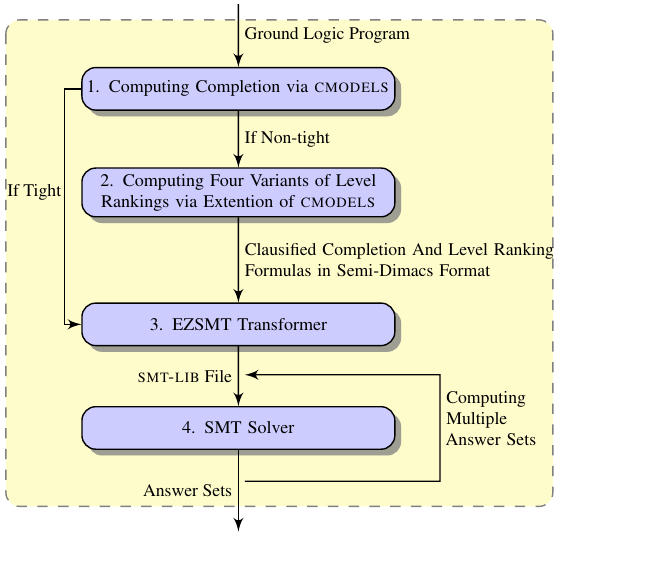}
\vspace{-2em}
\captionof{figure}{\diff Architecture}
\vspace{-2em}
\label{fig:syspipeline}
\end{center}
\end{figure}

Figure~\ref{fig:syspipeline} illustrates the pipeline architecture of \hbox{\diff}
\footnote{\diff is posted at { https://www.unomaha.edu/college-of-information-science-and-technology/natural-language-processing-and-knowledge-representation-lab/software/cmodels-diff.php} }. 
It is an extension of the \cmodels~\cite{lierphd} system.  
The \diff system takes an arbitrary (tight or non-tight) logic program in the language supported by {\cmodels} as an input. 
These logic programs may contain such features as choice rules and aggregate expressions. The rules with these features are translated away by {\cmodels}' original algorithms~\cite{lierphd}. 
Our main contribution is block 2, where the \diff system adds the corresponding
level ranking formula if the program is not tight. After that, the transformer taken from \ezsmt v1.1 is called to convert output from block 2 into \smtlib syntax. Finally, any SMT solvers supporting \smtlib, such as \cvcFour~\cite{cvc4url}, \zThree~\cite{z3url} and \yices~\cite{yicesurl}, can be called to compute solutions (that correspond to answer sets). 

%Currently, SMT solvers typically find only a single model. We also extend the architecture of \diff so that it allows computation of multiple solutions.
The \diff system allows us to compute multiple answer sets. Currently, SMT solvers typically find only a single model. We design a process to enumerate all models. After computing an answer set $X$ of a program, the \diff system invokes an SMT solver again by adding formulas encoding the fact that a newly computed model should be different from~$X$. This process is repeated until the pre-specified number of solutions is enumerated or it has been established that no more solutions exist.

\vspace{-0.1em}
\subsection{SMT-based Constraint Answer Set solver \ezsmtPlus}
\vspace{-0.1em}

\begin{figure}[t!]

\begin{center}
\captionsetup{type=figure}
\includegraphics[width=90mm]{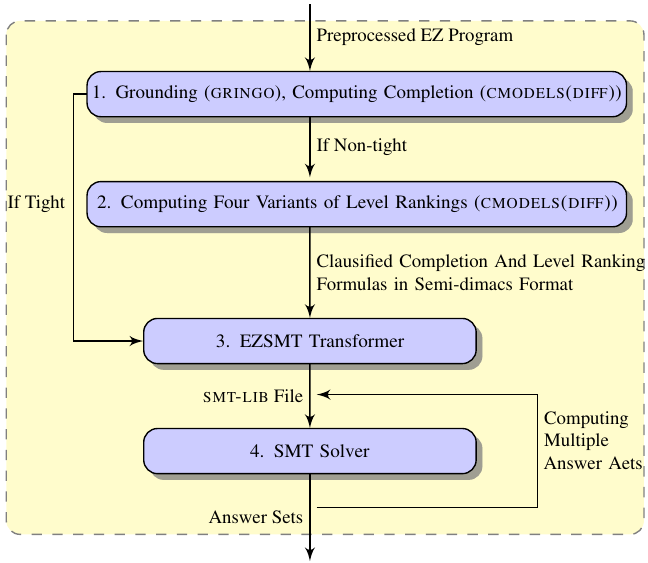}
\captionof{figure}{\ezsmtPlus Architecture}
\label{fig:syspipeline2}
\vspace{-2em}
\end{center}
\end{figure}

\bigbreak
We utilize the \diff system to build the extension \ezsmtPlus \footnote{\ezsmtPlus is available at { https://www.unomaha.edu/college-of-information-science-and-technology/natural-language-processing-and-knowledge-representation-lab/software/ezsmt.php} }. Figure~\ref{fig:syspipeline2} illustrates its architecture. The \ezsmtPlus system accepts an arbitrary (\textit{tight} or \textit{non-tight}) CASP program as input, and utilizes existing ASP tools for grounding and computing \textit{completion}~\cite{fag94}. Our main contribution is block 2, which produces corresponding formulas for a non-tight program. These formulas are a combination of the program's \textit{completion} 
and \textit{level ranking} formulas. We find a way to set minimal upper bounds for \textit{level rankings} in order to reduce search space. 
Then, the original \ezsmt transformer is used to translate them into a text file written in \smtlib, and an SMT solver is called to find models of these formulas. The procedure used by \ezsmtPlus to compute multiple solutions is identical to that of \diff. 

\paragraph{Experiments} We benchmark \ezsmtPlus with eight problems. 
%\diff is compared to ASP solvers \clingo and \cmodels.
\ezsmtPlus is compared to state-of-the-art CASP solvers \clingcon~\cite{clingconurl} and \ezcsp~\cite{ezcspconurl}. The experimental results are presented in the paper~\cite{shen182}.
They show that \ezsmtPlus is a viable tool for finding \textit{answer sets} of CASP programs, and can solve some difficult instances where its peers time out. Utilizing different SMT solvers may improve the performance of \ezsmtPlus in the future.

 \paragraph{Significance} The \ezsmtPlus system removes the inability to process non-tight input and enumerate multiple answer sets. It provides new capabilities towards utilizing declarative answer set programming paradigm for problems containing a wide variety of constraints including linear constraints over real numbers, mixed integer and real numbers, as well as nonlinear constraints. We believe that, by making clear the translation of arbitrary CASP logic programs to SMT, our work will boost the cross-fertilization between the two areas.

\vspace{-0.2em}
\section{Future Work}
\vspace{-0.2em}

In the future, we will extend \ezsmtPlus to allow processing of optimization statements. Using partial weighted maxSMT problems \cite{Fu06} is one potential approach. Yet, it requires theoretical work on connecting semantics of ASP and SMT constructs for optimizations.

The technique implemented by our systems for enumerating multiple answer sets of a program is basic. In the future we would like to adopt a nontrivial methods for model enumeration discussed in \cite{geb07c} to our settings.

The contributions of our work also open a door to the development of a novel constraint-based method in processing logic programs by producing intermediate output in  {\sc minizinc}~\cite{minizinc} in place of \smtlib.

\bibliographystyle{splncs04}
\bibliography{abstractmods-bib}
\end{document}